\documentclass[manuscript,]{aastex631}

\usepackage{url}
\usepackage{graphicx}
\usepackage{float}
\usepackage{amstext,amsmath,amssymb,amsthm}
\usepackage{upgreek}
\usepackage{enumerate}
\usepackage{txfonts}
\usepackage{graphicx}
\usepackage{subfigure}
\usepackage{multirow}
\usepackage[T1]{fontenc}
\let\tablenum\relax
\usepackage{siunitx}
\usepackage{color}
\usepackage{xcolor}
\usepackage{ulem}

\shorttitle{Insight-HXMT Timing calibration}

\begin{document}

\title{In-orbit timing calibration of \textit{Insight-Hard X-ray Modulation Telescope}}

\correspondingauthor{Xiaobo Li}
\email{lixb@ihep.ac.cn}


\author[0000-0003-3127-0110]{Youli Tuo}
\affiliation{Key Laboratory of Particle Astrophysics, Institute of High Energy Physics, Chinese Academy of Science, Beijing 100049, Peoples's Republic of China}

\author[0000-0003-4585-589X]{Xiaobo Li}
\affiliation{Key Laboratory of Particle Astrophysics, Institute of High Energy Physics, Chinese Academy of Science, Beijing 100049, Peoples's Republic of China}

\author[0000-0002-3776-4536]{Mingyu Ge}
\affiliation{Key Laboratory of Particle Astrophysics, Institute of High Energy Physics, Chinese Academy of Science, Beijing 100049, Peoples's Republic of China}

\author{Jianyin Nie}
\affiliation{Key Laboratory of Particle Astrophysics, Institute of High Energy Physics, Chinese Academy of Science, Beijing 100049, Peoples's Republic of China}

\author[0000-0003-0274-3396]{Liming Song}
\affiliation{Key Laboratory of Particle Astrophysics, Institute of High Energy Physics, Chinese Academy of Science, Beijing 100049, Peoples's Republic of China}
\affiliation{University of Chinese Academy of Sciences, Chinese Academy of Sciences, Beijing, People's Republic of China.}

\author[0000-0002-8476-9217]{Yupeng Xu}
\affiliation{Key Laboratory of Particle Astrophysics, Institute of High Energy Physics, Chinese Academy of Science, Beijing 100049, Peoples's Republic of China}
\affiliation{University of Chinese Academy of Sciences, Chinese Academy of Sciences, Beijing, People's Republic of China.}

\author[0000-0003-2256-6286]{Shijie Zheng}
\affiliation{Key Laboratory of Particle Astrophysics, Institute of High Energy Physics, Chinese Academy of Science, Beijing 100049, Peoples's Republic of China}

\author[0000-0003-3248-6087]{Fangjun Lu}
\affiliation{Key Laboratory of Particle Astrophysics, Institute of High Energy Physics, Chinese Academy of Science, Beijing 100049, Peoples's Republic of China}
\affil{Key Laboratory of Stellar and Interstellar Physics and School of Physics and Optoelectronics, Xiangtan University, Xiangtan 411105, Hunan, China}

\author[0000-0001-5586-1017]{Shuang-Nan Zhang}
\affiliation{Key Laboratory of Particle Astrophysics, Institute of High Energy Physics, Chinese Academy of Science, Beijing 100049, Peoples's Republic of China}
\affiliation{University of Chinese Academy of Sciences, Chinese Academy of Sciences, Beijing, People's Republic of China.}

\author[0000-0002-4834-9637]{Congzhan Liu}
\affiliation{Key Laboratory of Particle Astrophysics, Institute of High Energy Physics, Chinese Academy of Science, Beijing 100049, Peoples's Republic of China}

\author{Xuelei Cao}
\affiliation{Key Laboratory of Particle Astrophysics, Institute of High Energy Physics, Chinese Academy of Science, Beijing 100049, Peoples's Republic of China}

\author{Yong Chen}
\affiliation{Key Laboratory of Particle Astrophysics, Institute of High Energy Physics, Chinese Academy of Science, Beijing 100049, Peoples's Republic of China}

\author{Jinlu Qu}
\affiliation{Key Laboratory of Particle Astrophysics, Institute of High Energy Physics, Chinese Academy of Science, Beijing 100049, Peoples's Republic of China}

\author{Shu Zhang}
\affiliation{Key Laboratory of Particle Astrophysics, Institute of High Energy Physics, Chinese Academy of Science, Beijing 100049, Peoples's Republic of China}

\author{Haisheng Zhao}
\affiliation{Key Laboratory of Particle Astrophysics, Institute of High Energy Physics, Chinese Academy of Science, Beijing 100049, Peoples's Republic of China}

\author{Shuo Xiao}
\affiliation{Key Laboratory of Particle Astrophysics, Institute of High Energy Physics, Chinese Academy of Science, Beijing 100049, Peoples's Republic of China}
\affiliation{University of Chinese Academy of Sciences, Chinese Academy of Sciences, Beijing, People's Republic of China.}

\author{Baiyang Wu}
\affiliation{Key Laboratory of Particle Astrophysics, Institute of High Energy Physics, Chinese Academy of Science, Beijing 100049, Peoples's Republic of China}
\affiliation{University of Chinese Academy of Sciences, Chinese Academy of Sciences, Beijing, People's Republic of China.}

\author{Xiangyang Wen}
\affiliation{Key Laboratory of Particle Astrophysics, Institute of High Energy Physics, Chinese Academy of Science, Beijing 100049, Peoples's Republic of China}

\author{Weichun Jiang}
\affiliation{Key Laboratory of Particle Astrophysics, Institute of High Energy Physics, Chinese Academy of Science, Beijing 100049, Peoples's Republic of China}

\author{Bin Meng}
\affiliation{Key Laboratory of Particle Astrophysics, Institute of High Energy Physics, Chinese Academy of Science, Beijing 100049, Peoples's Republic of China}

\author{Weiwei Cui}
\affiliation{Key Laboratory of Particle Astrophysics, Institute of High Energy Physics, Chinese Academy of Science, Beijing 100049, Peoples's Republic of China}

\author{Wei Li}
\affiliation{Key Laboratory of Particle Astrophysics, Institute of High Energy Physics, Chinese Academy of Science, Beijing 100049, Peoples's Republic of China}

\author{Yifei Zhang}
\affiliation{Key Laboratory of Particle Astrophysics, Institute of High Energy Physics, Chinese Academy of Science, Beijing 100049, Peoples's Republic of China}

\author{Xufang Li}
\affiliation{Key Laboratory of Particle Astrophysics, Institute of High Energy Physics, Chinese Academy of Science, Beijing 100049, Peoples's Republic of China}

\author{Yanji Yang}
\affiliation{Key Laboratory of Particle Astrophysics, Institute of High Energy Physics, Chinese Academy of Science, Beijing 100049, Peoples's Republic of China}

\author{Ying Tan}
\affiliation{Key Laboratory of Particle Astrophysics, Institute of High Energy Physics, Chinese Academy of Science, Beijing 100049, Peoples's Republic of China}

\author[0000-0002-0238-834X]{Bing Li}
\affiliation{Key Laboratory of Particle Astrophysics, Institute of High Energy Physics, Chinese Academy of Science, Beijing 100049, Peoples's Republic of China}





\begin{abstract}

We describe the timing system and the timing calibration results of the three payloads on-board the \textit{Insight-Hard X-ray Modulation Telescope} (\textit{Insight-HXMT}). These three payloads are the High Energy X-ray telescope (HE, 20-250\,keV), the Medium Energy X-ray telescope (ME, 5-30\,keV) and the low Energy X-ray telescope (LE, 1-10\,keV). We present a method to correct the temperature-dependent period response and the long-term variation of the on-board crystal oscillator, especially for ME that does not carry a temperature-compensated crystal oscillator. 
The time of arrivals (ToAs) of the Crab pulsar are measured to evaluate the accuracy of the timing system. As the ephemeris of the Crab pulsar given by Jodrell Bank observatory has systematic errors around \citep{2004Rots} \SI{40}{\micro\second}, we use the quasi-simultaneous observations of the X-ray Timing Instrument (XTI) on-board the Neutron star Interior Composition Explorer (\textit{NICER}) to produce the Crab ephemerides and to verify the timing system of \textit{Insight-HXMT}.
The energy-dependent ToAs' offsets relative to the \textit{NICER} measurements including physical and instrumental origins are about \SI{24.7}{\micro\second}, \SI{10.1}{\micro\second} and \SI{864.7}{\micro\second}, and the systematic errors of the timing system are determined as \SI{12.1}{\micro\second}, \SI{8.6}{\micro\second}, and \SI{15.8}{\micro\second}, for HE, ME and LE respectively.

\end{abstract}



\section{Introduction}
\label{sec:1 introduction}
The \textit{Hard X-ray Modulation Telescope}, dubbed as \textit{Insight-HXMT}, was successfully launched on June 15th, 2017 into a low earth orbit with an altitude of 550\,km and an inclination of 43\,degrees \citep{Zsn2019, LuFangjun..HXMT..2018, 2020JHEAp..27...64L}. There are three payloads on-board \textit{Insight-HXMT}, the High Energy X-ray telescope (HE) that contains 18 NaI(Tl)/CsI(Na) phoswich scintillation detectors for observations in 20--250\,keV \citep{Lcz2019}, the Medium Energy X-ray telescope (ME) with 1728 Si-PIN detectors for 5--30\,keV \citep{Cxl2019}, and the Low Energy X-ray telescope (LE) with 96 SCD detectors for 1--15\,keV \citep{Cy2019}.


Fast X-ray timing observation of astrophysical objects is one of the key capabilities and tasks of \textit{Insight-HXMT}. High timing accuracy is essential to achieve the scientific goals involving rapid variability, such as rotation and accretion-powered millisecond pulsars, gamma ray bursts and kHz quasi-periodic oscillations. This is the reason that \textit{Insight-HXMT} carries a GPS receiver\,(GPSR) on-board. The high precision of the \textit{Insight-HXMT} timing system was at least partly demonstrated in the experiments of autonomous space navigation with pulsars \citep{2019ApJS..244....1Z}, but an exact calibration has not yet been performed. In this paper, the main purpose is to perform an in-orbit timing calibration for \textit{Insight-HXMT}.
Most missions in X-ray and Gamma-ray perform in-orbit timing calibration by observing pulsars, especially millisecond pulsars, which have stable timing properties. However, millisecond pulsars are very faint for \textit{Insight-HXMT} due to its high background level \citep{2020JHEAp..27...14L,2020JHEAp..27...44G,2020JHEAp..27...24L} and thus the Crab pulsar is the most suitable source for timing calibration of \textit{Insight-HXMT}.

Actually the timing calibrations of  some other X-ray or Gamma-ray telescopes are also carried out based on the observations of the Crab pulsar. The timing calibration, based on the Crab pulsar in X-ray and Gamma-ray bands, typically measures the delay of the radio main pulse of the Crab pulsar compared to the main pulse observed by the respective satellite. \cite{2021Sci...372..187E} measured the difference between Crab's X-ray and radio pulse profiles and radio observations (using radio telescopes at Usuda and Kashima),
and found that the main pulse ahead of radio main pulse and deviation of \textit{NICER} are about \SI{304}{\micro\second} and \SI{1.3}{\micro\second};
the main pulse observed by 
\textit{Suzaku}/HXD precedes \textit{RXTE}/PCA,  \textit{RXTE}/HEXTE, \textit{INTEGRAL}/IBIS ISGRI, and \textit{Swift}/BAT by $0.003\pm 0.001$, $0.002\pm0.001$, $0.002\pm0.001$, $0.002\pm0.001$ periods, respectively \citep{terada2008orbit}.
Thus the offset of \textit{Suzaku}/HXD leads those X-ray instruments by about \SI{70}{\micro\second},
and the systematic error is \SI{270}{\micro\second} obtained by Jodrell Bank ephemeris. The following instruments verify the offsets and systematic errors based on the ephemerides given by Jodrell Bank observatory as well. \textit{RXTE}/PCA's offset to radio Jodrell Bank observation is \SI{344}{\micro\second} and systematic error is \SI{40}{\micro\second} \citep{2004Rots}; \textit{INTEGRAL}/SPI offers offset of \SI{280}{\micro\second} and systematic error of \SI{40}{\micro\second} \citep{2003A&A...411L..31K,molkov2010absolute}; \textit{Fermi}/GBM's 
offset is $222\pm 4$\,\si{\micro\second}	and systematic error is \SI{56}{\micro\second}, while \textit{Fermi}/LAT's 
offset is $111 \pm 4$\,\si{\micro\second} and systematic error of \SI{57}{\micro\second} \citep{aliu2011detection}, and all results are epoch-folded by Jodrell Bank ephemeris. 
For \textit{RXTE}/PCA and \textit{INTEGRAL}/SPI, the systematic error obtained by Jodrell Bank ephemeris is mainly caused by the interstellar scattering and instrument calibration of the radio telescope \citep{2004Rots}. Therefore, we decide to utilize the observations of \textit{NICER} to produce the Crab ephemeris and to estimate the offset and systematic error of \textit{Insight-HXMT} in the quasi-simultaneous time intervals.

In this paper, we give a description of the timing system of \textit{Insight-HXMT}. The detailed time assignment procedure and the correction on the period of crystal oscillator are presented in Section \ref{sec:2 instrument description}. In section \ref{sec:crabresults}, we describe our measurements, data processing and timing analysis method on the Crab pulsar observations. In section \ref{sec:timingresults}, the Crab ephemeris are derived by \textit{NICER}/XTI observations. The offsets relative to \textit{NICER}/XTI measurements and the systematic errors of \textit{Insight-HXMT} timing system are presented by the ephemerides. The Section \ref{sec:conclusion} summarizes our results.

\section{The timing system of \textit{Insight-HXMT}} \label{sec:2 instrument description}
\textit{Insight-HXMT} satellite carries a GPSR to obtain accurate time information from GPS satellites and distributes the GPS pulse per second (PPS) to each payload. After about 250\,ms, the mission elapsed time (MET) for the GPS pulse, relative to 2012-01-01T00:00:00 (UTC), is also sent to each payload. The GPS pulse and MET time are stored as a pair in the raw data package of each payload. If the GPS fails or is locked off, an alternative module in GPSR will generate PPS.

\begin{figure}
	\centering
	\includegraphics[width=1.0\textwidth]{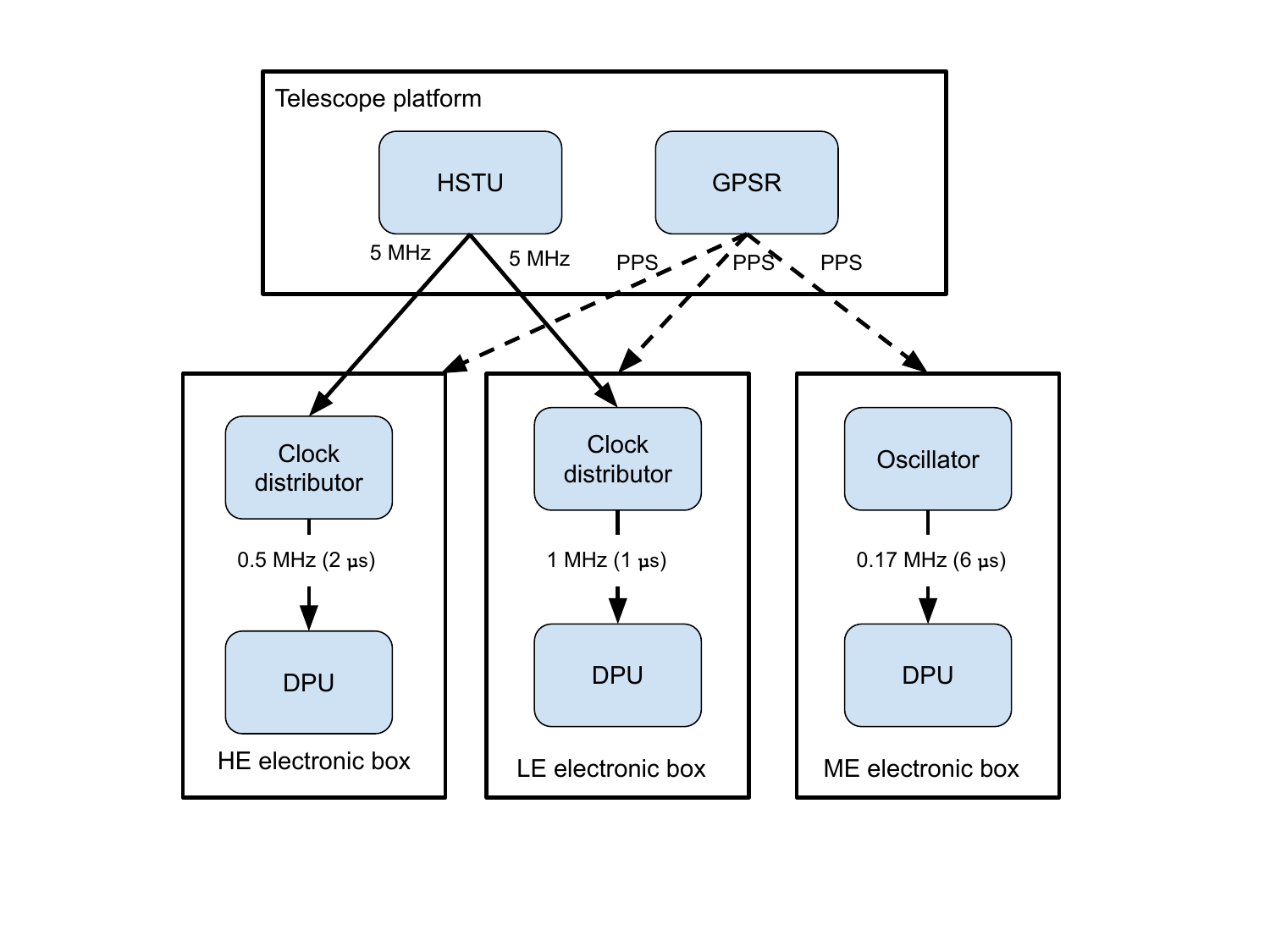}
	\caption{The schematic diagram of the timing system of \textit{Insight-HXMT}. GPS receiver (GPSR) sends GPS pulse per second (PPS) to the electronic boxes of the three payloads. The high-stability time unit (HSTU) on the platform distributes 5\,MHz clocks to HE and LE. The clock distributor in HE uses the 5\,MHz clocks to generate local clock with period of \SI{2}{\micro\second} for the data processing unit (DPU) of HE. The clock distributor in LE also utilizes the 5\,MHz clock to generate the local clock with period of \SI{1}{\micro\second} and sends to DPU. ME uses a quartz oscillator instead to generate \SI{6}{\micro\second} clock to DPU.}
	\label{FIG:GPSR}
\end{figure}

\textit{Insight-HXMT} also carries a high-stability time unit (HSTU) with a frequency of 5\,MHz and stability of $3\times 10^{-8}\,\rm{s}^{-1}$. The clock distributor in the HE and LE payloads use this 5\,MHz to generate \SI{2}{\micro\second} and \SI{1}{\micro\second} local clock for the data processing unit (DPU), which mean that the time of  the HE events is in units of \SI{2}{\micro\second} and the LE events \SI{1}{\micro\second}. ME carries an internal quartz oscillator to generate the \SI{6}{\micro\second} local clock by itself. Figure \ref{FIG:GPSR} shows a schematic diagram of the timing system of \textit{Insight-HXMT}.

\subsection{Time assignment of \textit{Insight-HXMT} events} \label{sec:2.1 HE instrument}
In this subsection, we take HE as an example to show the time assignment of the physical events recorded by \textit{Insight-HXMT}. The timing information of HE events is divided into two parts, one is the integral part of the second (defined as Stime hereafter) and the other is the fractional part of the second (defined as Ptime hereafter).
Stime is produced using the GPS pulse provided by GPSR, which is inserted into the physical data stream per second with a special tag.
Ptime is the local time count of the \SI{2}{\micro\second} clock cycle.
In the raw package of HE, Stime has 32 bits and is increased by one once HE receives the GPS pulse. Therefore, Stime covers time ranges from zero to about 136 years ($2^{32}\,\rm{s}$) if the payload is not turned off and restarted. Ptime has 19 bits and records the local count of clock with cycles of \SI{2}{\micro\second}. As a result, Ptime counter will overflow and re-count from zero about every \SI{1.048}{\second} ($2^{19}\times$\SI{2}{\micro\second}). 

In order to compress the amount of ME and LE data, only the lowest 14 bits of Ptime are recorded for each event in ME. A carry-up event is generated when the Ptime counter overflows and resets to zero about every 0.098s ($2^{14}\times$\SI{6}{\micro\second}). The number of the carry-up event is recorded in the raw data with 32 bits length. LE also records the lowest 16 bits of Ptime for each event, and a carry-up event is also generated when the Ptime counter overflows and recounts from zero about every 0.065s ($2^{16}\times$\SI{1}{\micro\second}). The number of the carry-up event is also recorded in the raw data with 32 bits length.
As listed in Table \ref{tab:Ptime}, the local time counters of Ptime are summarized for the three payloads on-board \textit{Insight-HXMT}. The period of the local clock is also the time resolution of the three payloads.

\begin{table}[htb]\footnotesize
  \centering
  \caption{The local time counters of Ptime for \textit{Insight-HXMT} payloads.}
  \label{tab:Ptime}
   \begin{tabular}{ccc}
    \hline
    Payloads & Bit length  & Period of the local clock \\
        \hline
     HE & 19 bits & \SI{2}{\micro\second}   \\
    ME & 14 bits & \SI{6}{\micro\second}   \\
    LE & 16 bits & \SI{1}{\micro\second}   \\
    \hline
    \end{tabular}
\end{table}

\begin{figure}
	\centering
		\includegraphics[width=1.0\textwidth]{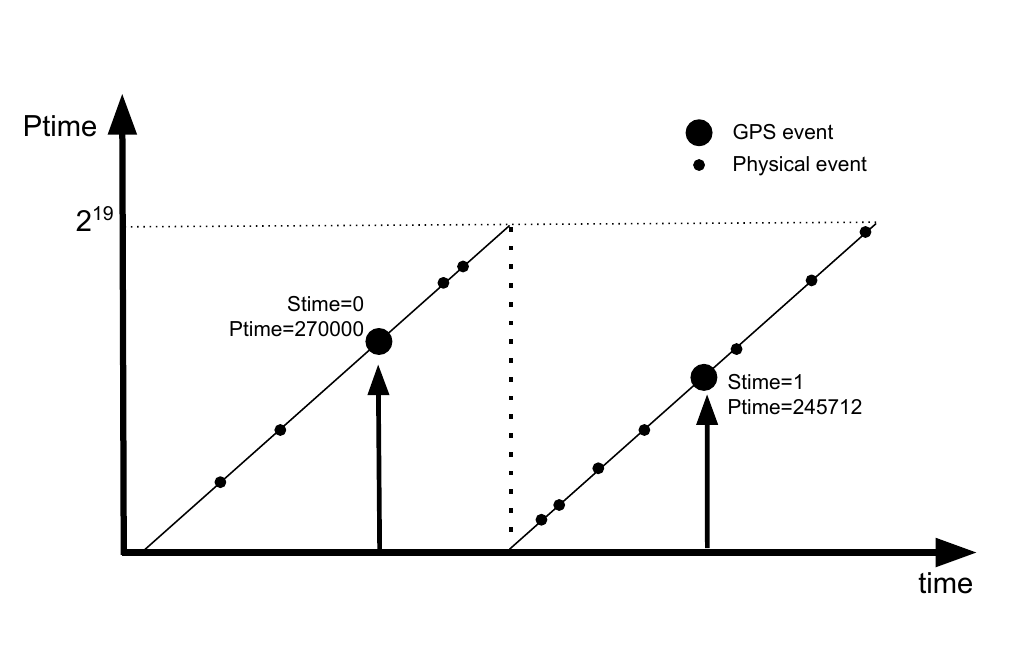}
	\caption{The time assignment of \textit{Insight-HXMT}/HE events. GPS events recorded by Stime and Ptime are shown with big solid circle and inserted into the event stream every second. The physical events with small solid circle are only recorded by Ptime. Ptime accumulates to $2^{19}$ and starts counting from zero.}
	\label{FIG:HEevents}
\end{figure}

The arrival times of X-ray photons from astronomical objects and particles from background are recorded by Ptime only. The GPS pulses are recorded by Stime and Ptime simultaneously as displayed in Figure \ref{FIG:HEevents}. The physical events are filled between GPS events, and the rate of physical events is decided by the intensities of the astronomical objects and background.
Since HSTU is not synchronized to GPSR on-board, the local clock should be corrected by GPS pulses with offline software.

\subsection{Correction of the arrival times of events} \label{sec:2.2 ME instrument}
We use an automated data production software to decode the data in the raw package and save the data files in the flexible image transport system (FITS) format. The decoding process calculates the physical values, such as time, pulse height, pulse width, and writes them to the output FITS files.

The temperature-dependent and long-term variation effects on the period of the local clock are corrected by synchronizing Ptime and PPS. In each hour, the Stime of the first GPS pulse can be labeled as $S_{0}$. The nominal local period is \SI{2}{\micro\second} for Ptime, but there may exist jitters. Therefore, the actual period, first and second order derivative of the period in each hour are represented by $\tau$, $\dot{\tau}$ and $\ddot{\tau}$, respectively.
The Stime of the GPS pulse, labeled as $S$, increases by one per second and can be modeled as the following Taylor expansion,
\begin{equation}\label{equ:timecal}
   S = S_{0}+ \tau(N-N_{0})+\frac{1}{2}\dot{\tau}(N-N_{0})^{2}+\frac{1}{6}\ddot{\tau}(N-N_{0})^{3}+...,
\end{equation}
where $N_{0}$ is the Ptime of the first GPS pulse in each hour, and $N$ is the accumulated local time counts of Ptime in reference to $N_{0}$.
The least square fit method is used to obtain the values of $S_{0}$, $\tau$, $\dot{\tau}$ and $\ddot{\tau}$.
Then $S$ of the physical events can be derived from Equation (\ref{equ:timecal}) according to the accumulated local time counts $N$ in reference to $N_{0}$ and the best fit parameters $S_0$, $\tau$, $\dot{\tau}$, $\ddot{\tau}$.

Since each GPS pulse corresponds to an MET time (labeled as $M$), so the MET time of events can be derived as,
 \begin{equation}\label{equ:timeMED}
    M = M_{0} + S - S_{0},
\end{equation}
where $M_{0}$ represents the MET time of the first GPS pulse in each hour. Then the MET time of an event can be converted into the terrestrial time (TT) value, which is the conventional time system used in the FITS file in X-ray astronomy.

\begin{figure}
	\centering
		\includegraphics[trim={3cm 8cm 3cm 8cm},clip, width=0.8\textwidth]{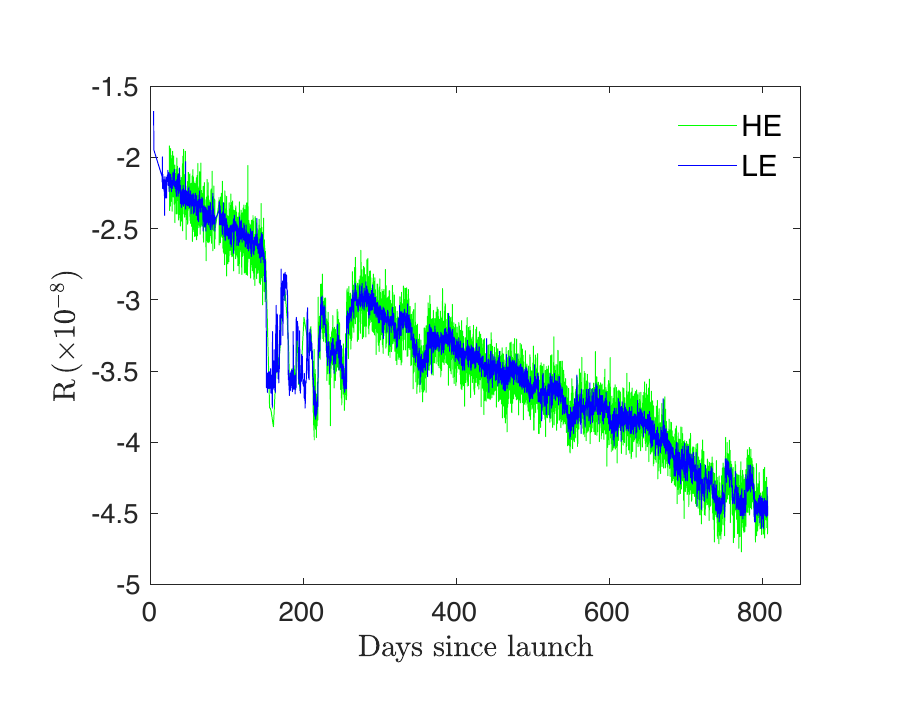}
	\caption{The evolution trends of the relative change of the nominal period for HE and LE. The blue line represents the result of LE and the green line represents the result of HE. Reasons for the deviations from the long time trend around day 200 are yet unknown.}
	\label{FIG:HEtimestep}
\end{figure}

We examined the reliability and accuracy of the time assignment process both before and after the launch. We found that $\tau$ and $\dot{\tau}$ are enough to describe $S$ for HE and LE. But for ME, $\ddot{\tau}$, and $\dddot{\tau}$ should be also included to describe the $S$. The nominal period ($\tau_{0}$) of the three payloads are \SI{2}{\micro\second}, \SI{6}{\micro\second}, and \SI{1}{\micro\second} for HE, ME, and LE respectively. The relative change ($R$) of $\tau$ can be described as,
\begin{equation}\label{equ:Rstep}
    R = \frac{\tau - \tau_{0}}{\tau_{0}},
\end{equation}
In Figure \ref{FIG:HEtimestep}, the evolution trends of $R$ for HE and LE are shown respectively. HE and LE both use HSTU provided by the satellite, so the evolution is almost the same. 

\begin{figure}[htb]
  \centering
  \subfigure[]{
  \label{subfig:randtemptotime}
     \includegraphics[width=0.48\textwidth]{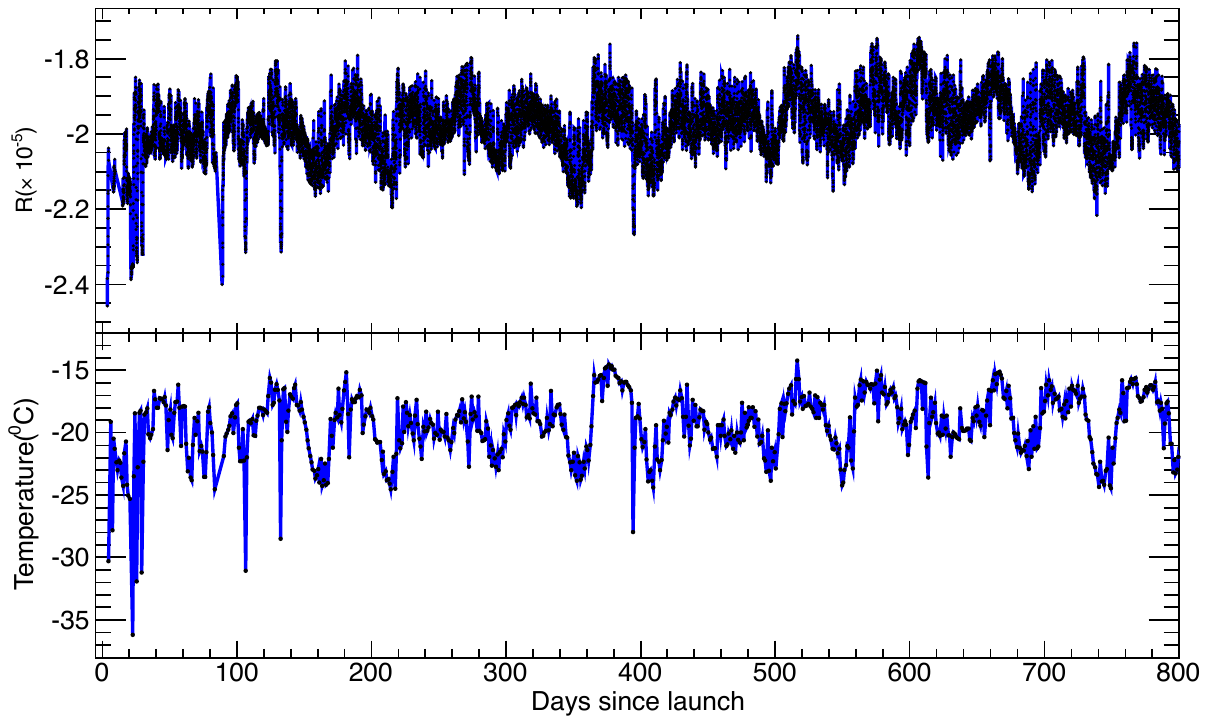}}
  \subfigure[]{
    \label{subfig:tempandr}
      \includegraphics[width=0.45\textwidth]{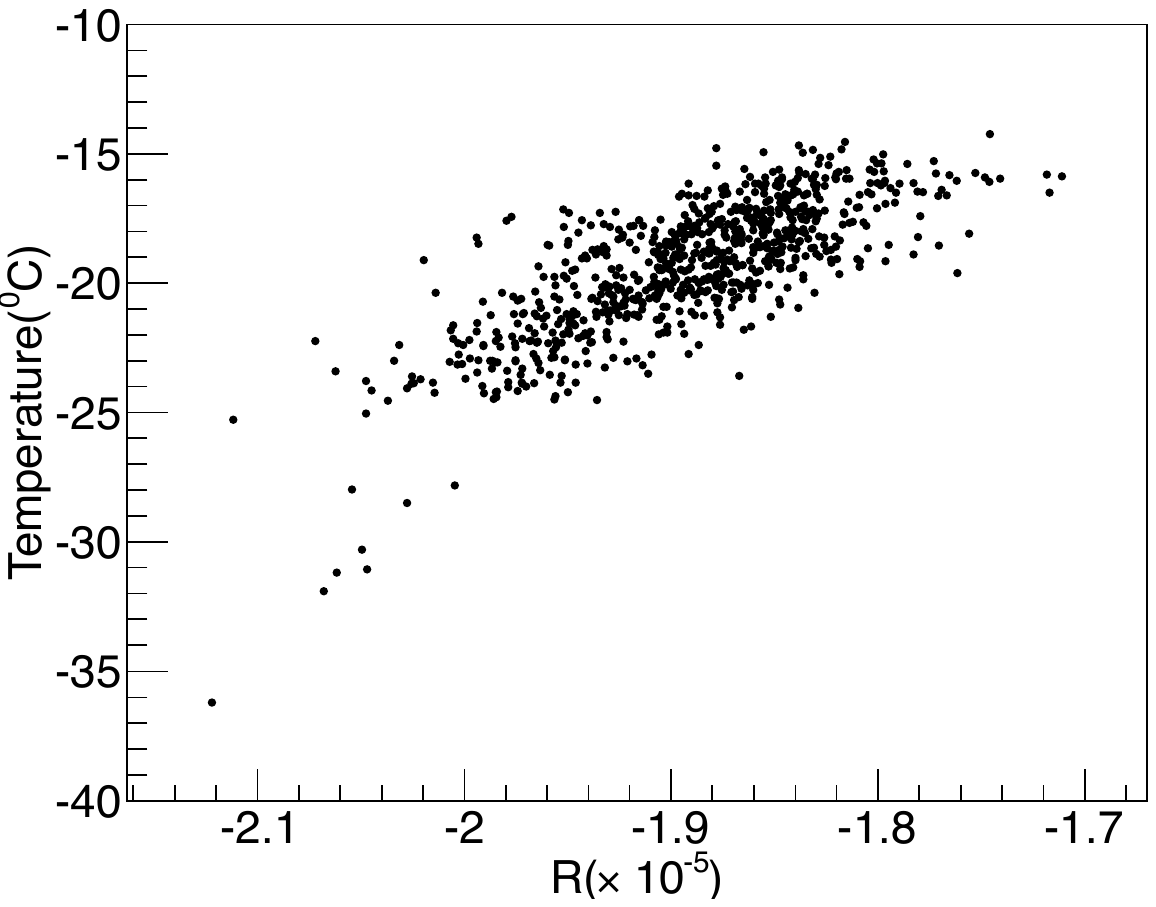}}
    \caption{(a) The top panel shows the evolution of the relative change of the nominal period for ME. The bottom panel shows the temperature versus days after launch. The real period of the quartz oscillation depends largely on temperature. (b) The correlation between temperature and relative change of the nominal period $R$.}
  \label{FIG:MEtimestep}
\end{figure}

Figure \ref{subfig:randtemptotime} shows the evolution of $R$ and temperatures measured by ME, and Figure \ref{subfig:tempandr} shows the correlation of $R$ and temperature. The period of the quartz oscillator depends largely on the temperature so they have a strong positive correlation. In other words, the effects of temperature variation on the quartz oscillator can be described well by Equation (\ref{equ:timecal}). As a result, using this method we can calibrate and correct the long-term variation of the period of the ME quartz oscillator and HSTU. Since ME uses the quartz oscillator carried by itself and the quartz oscillator is not temperature compensated, the relative change of the period supplied by this oscillator is three orders of magnitude higher than HSTU used by HE and LE.

\subsection{The stability of the local clock} \label{sec2.3}
\begin{figure}
	\centering
		\includegraphics[width=0.8\textwidth]{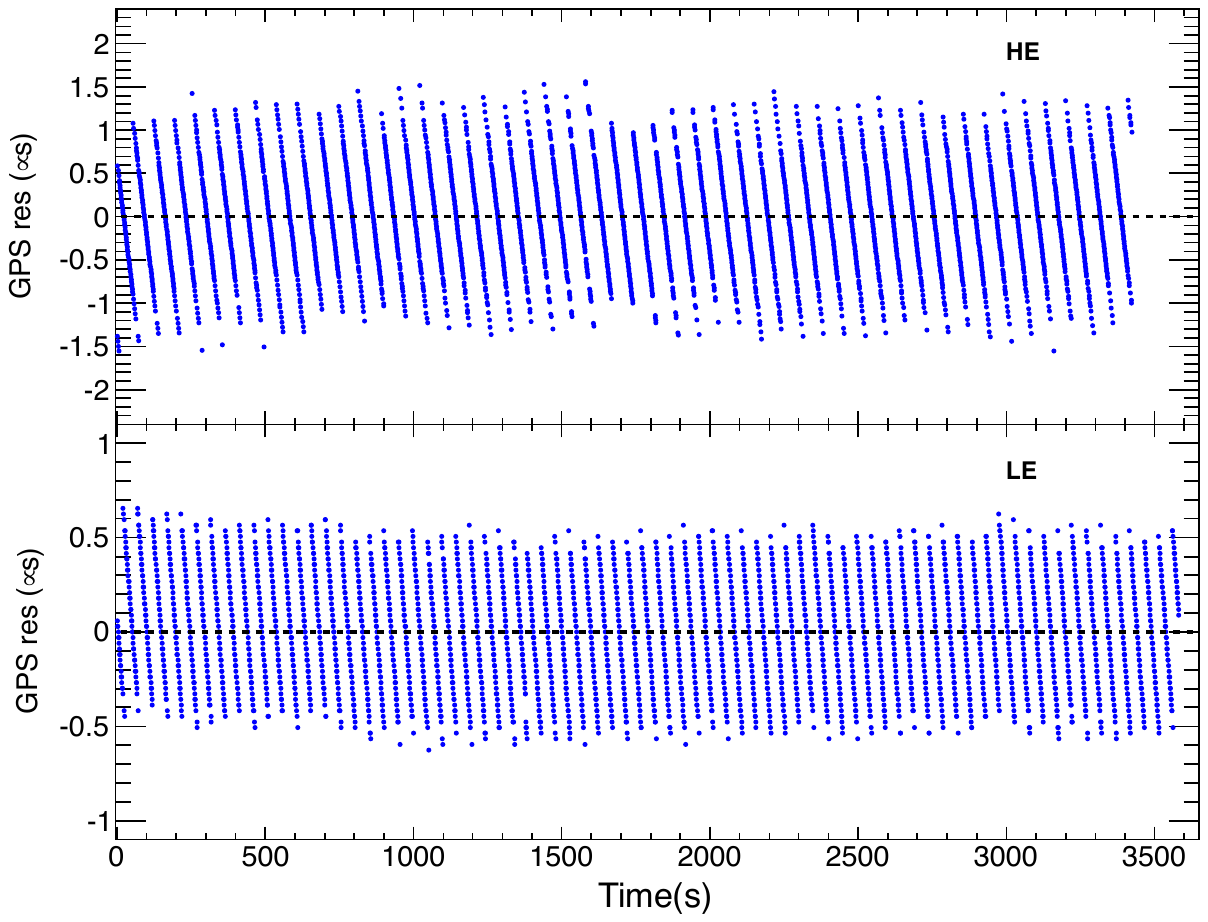}
	\caption{The residuals between the arrival time of GPS pulse in one hour calculated by Equation (\ref{equ:timecal}) and Stime which is an integer in seconds. The top panel shows the residual for HE, and the residual is less than \SI{2}{\micro\second} which is the period of the local clock. The bottom panel shows the residual for LE, and the residual is less than \SI{1}{\micro\second} which is the period of the local clock. The slope of each truncated line in each panel refers to the $R$ in Figure \ref{FIG:HEtimestep}.}
	\label{FIG:HELEGPSres}
\end{figure}

After the correction mentioned above, the stability of the local clock is revealed by the residuals between the recorded Stime of the GPS pulse and $S$ calculated by Equation (\ref{equ:timecal}).
We subtract Stime of the GPS pulse from the arrival time calculated by Equation (\ref{equ:timecal}). Their differences of HE and LE in one hour are shown in Figure \ref{FIG:HELEGPSres}. $S_{0}$, $\tau$ and $\dot{\tau}$ are used for HE and LE. The accuracy of the arrival time of GPS pulse is less than one period of the local clock for HE and LE.

\begin{figure}
	\centering
		\includegraphics[width=0.8\textwidth]{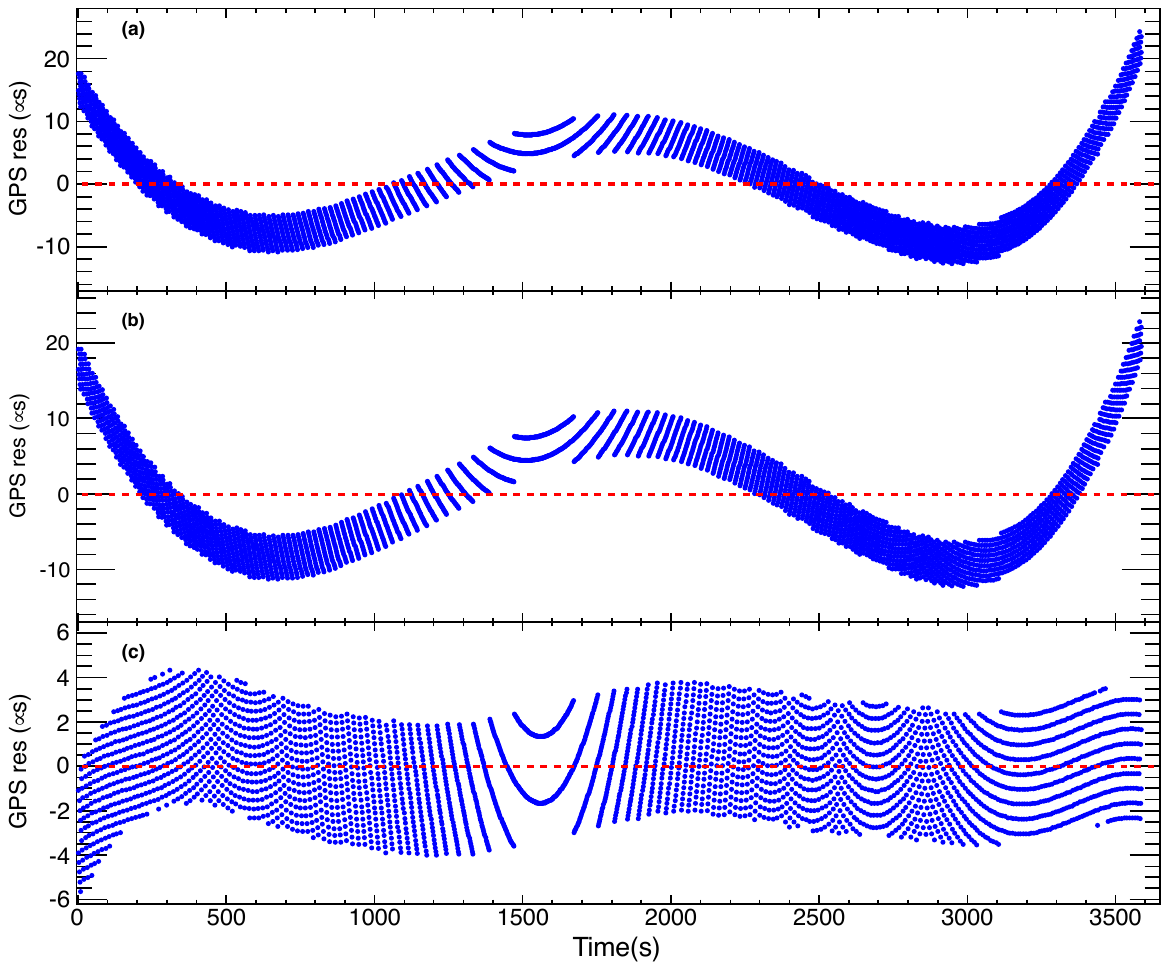}
	\caption{The residuals between the arrival time of GPS pulse in one hour calculated by Equation (\ref{equ:timecal}) and Stime for ME. Panel (a) shows the residual using $S_{0}$, $\tau$ and  $\dot{\tau}$. Panel (b) shows the residual using $S_{0}$ ,$\tau$ , $\dot{\tau}$ and $\ddot{\tau}$. Panel (c) shows the residual using $S_{0}$ ,$\tau$ ,$\dot{\tau}$, $\ddot{\tau}$ and $\dddot{\tau}$.}
	\label{FIG:MEGPSres}
\end{figure}

The differences between the arrival time of the GPS pulse calculated by Equation (\ref{equ:timecal}) and Stime recorded by ME are plotted in Figure \ref{FIG:MEGPSres}, in which panel (a) shows the results in one hour when only terms related to $S_{0}$, $\tau$ and  $\dot{\tau}$ are used in Equation (\ref{equ:timecal}). Panel (b) shows the results in one hour when we use $S_{0}$, $\tau$, $\dot{\tau}$, and $\ddot{\tau}$ in Equation (\ref{equ:timecal}). Panel (c) shows the results in one hour when $S_{0}$ ,$\tau$, $\dot{\tau}$, $\ddot{\tau}$ and $\dddot{\tau}$ are used in Equation (\ref{equ:timecal}). Obviously, the third order derivative of $\tau$ is needed to describe the period of local clock and make the arrival time of GPS pulse in less than one period of the local clock for ME. 

The stability of the local clock after the correction are within $\pm$\SI{1}{\micro\second}, $\pm$\SI{3}{\micro\second}, and  $\pm$\SI{0.5}{\micro\second} for HE, ME, and LE, respectively. We can then further confirm the performance of the timing system with the observations of the Crab pulsar.

\section{Observations of the Crab pulsar and data analysis}
\label{sec:crabresults}
\subsection{Data reduction}
The standard data reduction threads are performed using the analysis software \texttt{HXMTDASv2.04}. The recommended criteria for generating the good time intervals (GTIs) are\footnote{see HXMT manual for details http://hxmten.ihep.ac.cn/SoftDoc.jhtml} 
the intervals when (1) elevation angle > 10 degrees; (2) geomagnetic cut-off rigidities > 8 GeV; (3) satellite not in SAA and 300 seconds intervals near SAA; (4) pointing deviation to the source less than 0.04 degree.  The Crab is the main calibration target for \textit{Insight-HXMT}, which conducted a long-term monitoring program with a total effective exposure of $\sim 777\,\rm{ks}$ (78 times of observations until 2020-02-21). Observations with quasi-simultaneous ones by \textit{NICER} are listed in table \ref{tab:obs_id}.

\begin{table}[htb]\footnotesize
    \centering
    \begin{tabular}{l l l l l l }
       \hline
           \textbf{HXMT-Obs\_ID} & \textbf{Start (UTC)} & \textbf{Stop (UTC)} & \textbf{NICER-Obs\_ID} & \textbf{Start (UTC)} & \textbf{Stop (UTC)} \\
           \hline
P0111605001  &  2017-11-09T04:03:40  &  2017-11-10T00:55:03  &  ni1013010109  &  2017-11-09T10:17:39  &  2017-11-09T10:21:25  \\
P0111605002  &  2017-11-10T16:39:27  &  2017-11-11T00:47:15  &  ni1013010110  &  2017-11-10T17:22:06  &  2017-11-10T20:42:54  \\
P0111605003  &  2017-11-11T16:31:37  &  2017-11-12T00:39:35  &  ni1013010111  &  2017-11-11T10:27:10  &  2017-11-11T10:36:03  \\ 
P0111605004  &  2017-11-12T16:23:59  &  2017-11-13T00:32:07  &  ni1013010112  &  2017-11-12T14:16:03  &  2017-11-12T14:23:31  \\
P0111605005  &  2017-11-13T16:16:35  &  2017-11-14T00:24:54  &  ni1013010113  &  2017-11-13T13:26:45  &  2017-11-13T23:57:07  \\
P0111605006  &  2017-11-14T16:09:27  &  2017-11-15T00:17:59  &  ni1013010114  &  2017-11-14T00:13:49  &  2017-11-14T23:23:31  \\
P0111605007  &  2017-11-15T16:02:39  &  2017-11-16T00:11:27  &  ni1013010115  &  2017-11-15T00:38:16  &  2017-11-15T23:49:56  \\ 
P0111605008  &  2017-11-16T15:56:18  &  2017-11-16T20:43:06  &  ni1013010116  &  2017-11-16T01:21:04  &  2017-11-16T23:00:20  \\ 
P0111605009  &  2017-11-17T15:50:26  &  2017-11-17T20:53:02  &  ni1013010117  &  2017-11-17T00:30:30  &  2017-11-17T12:54:22  \\ 
P0111605010  &  2017-11-18T15:45:02  &  2017-11-18T20:41:24  &  ni1013010118  &  2017-11-17T14:24:10  &  2017-11-17T23:43:24  \\ 
P0111605011  &  2017-11-19T15:40:00  &  2017-11-19T20:27:01  &  ni1013010119  &  2017-11-18T01:14:29  &  2017-11-18T22:53:48  \\ 
P0111605012  &  2017-11-20T15:34:53  &  2017-11-20T19:17:12  &  ni1013010120  &  2017-11-19T00:24:55  &  2017-11-19T23:47:12  \\
P0111605013  &  2017-11-21T17:04:41  &  2017-11-21T22:06:55  &  ni1013010121  &  2017-11-20T01:14:37  &  2017-11-20T19:51:15  \\ 
             &                       &                       &  ni1011010301  &  2017-11-20T15:28:39  &  2017-11-20T16:58:08  \\ 

        \hline
    \end{tabular}
    \caption{The quasi-simultaneous observations of the Crab by \textit{Insight-HXMT} and \textit{NICER}}
    \label{tab:obs_id}
\end{table}

The arrival times of photons at \textit{Insight-HXMT} are converted to the arrival time at barycentric center. \texttt{HXMTDASv2.04} contains the module for barycentric correction, named \textit{hxbary}, in which the supported solar ephemerides are JPL-DE200 and JPL-DE405. Both solar ephemerides are used, JPL-DE200 is used when the timing results are generated by Radio Jodrell Bank ephemeris, while JPL-DE405 is used to produce the timing results using the ephemeris generated by \textit{NICER} observation (see section \ref{sec:timingresults}). The position accuracy of \textit{Insight-HXMT} is better than \SI{100}{\meter}, which introduces an uncertainty of less than \SI{0.3}{\micro\second} to the barycentric corrections.

The cleaned data of \textit{NICER}/XTI are also used for timing analysis. The barycentric correction for photons observed by \textit{NICER}/XTI is performed by \texttt{barycorr} in \texttt{FTOOL} package. In particular, we adopt the position of the Crab Pulsar for the barycentric correction from the values reported by Jodrell Bank \citep{1993MNRAS.265.1003L}: RA = $05^{\rm{h}}34^{\rm{m}}31^{\rm{s}}.973$, Dec = $22^{\circ}00'52''.069$ .

\subsection{Frequency search}
\label{subsec:search}
We apply the epoch-folding method \citep{1987A&A...180..275L} to search the frequencies of the Crab pulsar to characterize the accuracy of \textit{Insight-HXMT} on the measurement of the pulsar's spinning frequency. In this scenario, the data events corrected by solar ephemeris JPL-DE200 are prepared as input for frequency search. The frequency step for \textit{Insight-HXMT} data is fixed at $10^{-8}$\,Hz. The $\chi^2$ value is calculated, and the frequency corresponding to the maximum value of $\chi^2$ indicates the best frequency at the midpoint of the observation time span.

\subsection{Time of arrivals (ToAs) of the pulse}
\label{subsec:toas}
After the spin frequency is derived, we can obtain the pulse profile of the Crab pulsar.
Based on the pulse profile, the time of arrivals (ToAs) of the pulses can be obtained by calculating the arrival time of the main peak. The stability of the main peak position provides information about the accuracy of the timing system of \textit{Insight-HXMT} and the intrinsic information of the pulsar. 
The phase of each photon is calculated by the timing parameters of Crab's ephemeris:
\begin{equation}
    \label{eq:phi}
    \Phi = f_0(t-t_0) + \frac{1}{2}f_1(t-t_0)^2 + \frac{1}{6}f_3(t-t_0)^3,
\end{equation}
where $f_0$, $f_1$, $f_2$, $f_3$ are the frequency, frequency derivative, the second, and the third derivative of frequency. $t_0$ is the reference time of the timing parameters. The ToAs of the cumulative pulses are calculated by the arrival time of the main peak, ToA = $t_0 + P \cdot \Delta \phi $, where $\Delta\phi$ is the phase of the main peak, and $P=1/f_0$ is the period. 

\begin{figure}
    \centering
    \includegraphics[width=0.8\textwidth, angle=90]{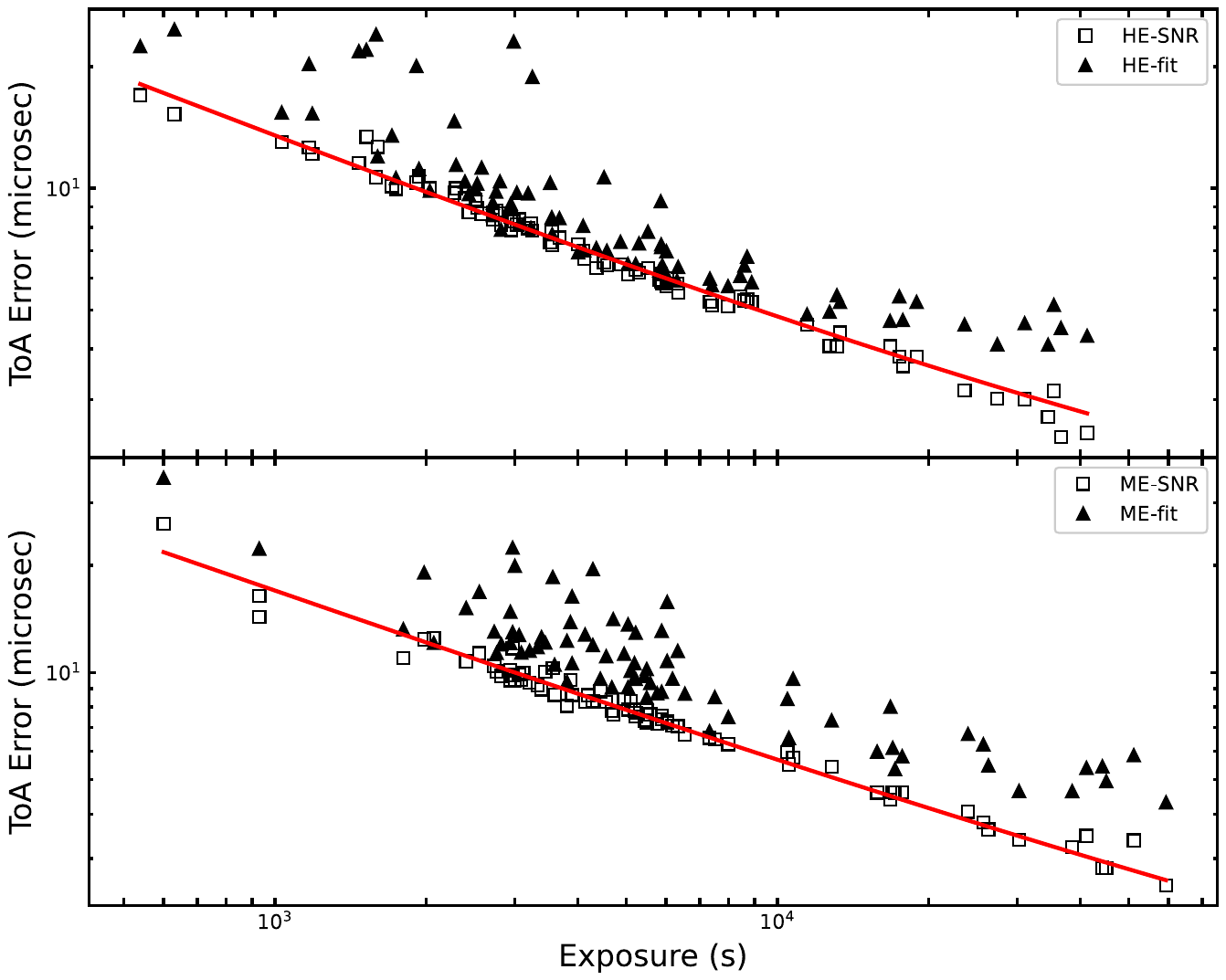}
    \caption{Top panel and the bottom panel show the relation between the ToA errors and the exposure time of corresponding profiles for HE, and ME, respectively. The errors of ToAs calculated from the signal-to-noise ratios of the main peak are presented as squares. And the errors the ToA fits are plotted as triangles. The red lines indicate the linear relation to the square root of the exposure time.}
    \label{fig:error-exp}
\end{figure}

A Gaussian function is used to fit the pulse profile within $\pm0.015$ of the maximum phase of the main peak. $\Delta\phi$ is the center value of the best-fit Gaussian function. The error of ToAs, $\sigma_{\rm{ToA}}$, can also be calculated from the signal-to-noise ratio of the main peak \citep{2019ApJ...887L..27G, 2012handbook}:
\begin{equation}
    \label{eq:toaerr}
    \sigma_{\rm{ToA}} = \frac{P\cdot \sigma_{\rm{Gauss}}}{N/\sqrt{N+N_{\rm{bkg}}}},
\end{equation}
where $\sigma_{\rm{Gauss}}$ is the standard deviation of the best-fit Gaussian function, $N$ is the event count of the source in the fitted phase range of the main peak, and $N_{\rm{bkg}}$ is the background event count in the main peak's phase range, determined by scaling the counts in the baseline phase range to the size of the main peak phase range. We also find that the error of ToAs calculated from the signal-to-noise ratio is equivalent to the error of the peak value in the Gaussian fitting. And they are both proportional to the square root of the exposure time, as shown in Figure \ref{fig:error-exp}.

\section{Timing results}
\label{sec:timingresults}

\subsection{Results of Jodrell Bank radio ephemeris for the Crab pulsar} \label{sec:Results}



\begin{figure}
	\centering
		\includegraphics[width=0.8\textwidth]{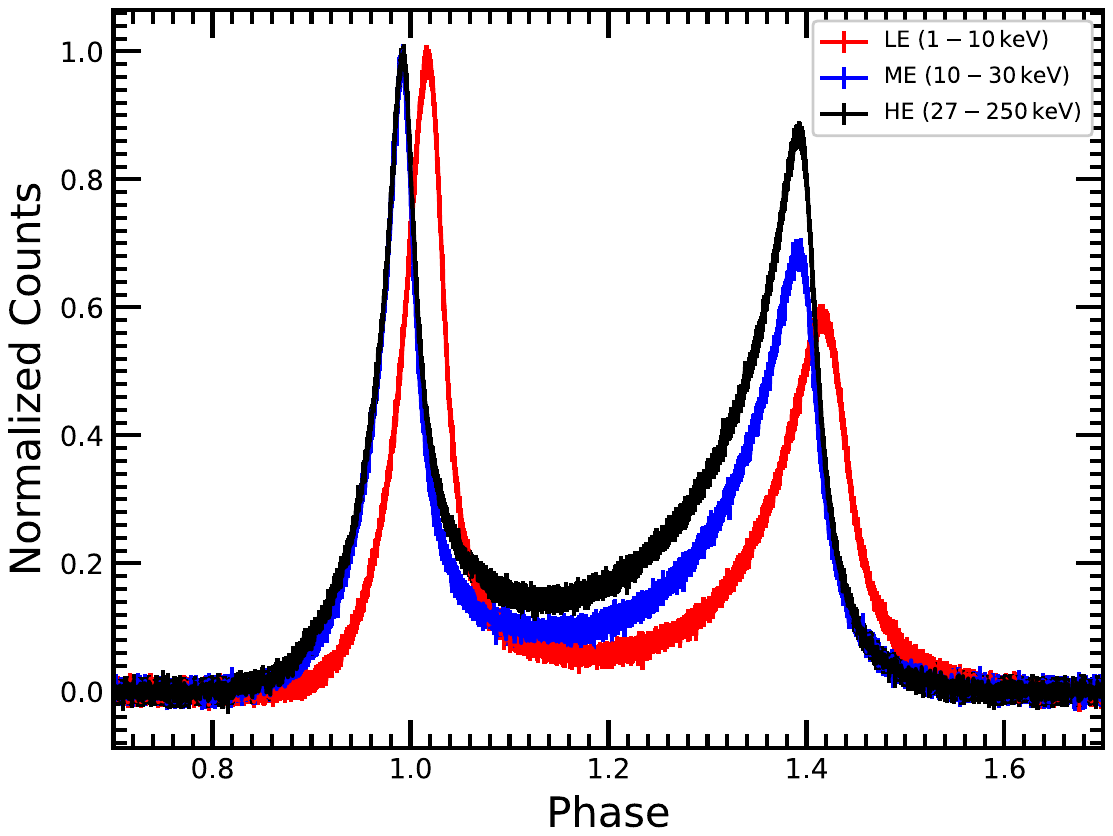}
	\caption{The pulse profile of the Crab pulsar observed by \textit{Insight-HXMT} in three energy bands, using the Jodrell Bank Crab ephemeris. The red, blue, and black lines indicate the profiles of the three payloads of \textit{Insight-HXMT}, LE (1--10\,keV), ME (10--30\,keV), and HE (27--250\,keV), respectively.  For clarify the phase is plotted from 0.7 to 1.7. The counts are normalized by scaling the counts of the main peak to unity and shifting the mean value of the background phase (0.6--0.8) to zero. The LE profile is significantly delayed by LE's readout mechanism (see text for details).}
	\label{FIG:CrabProfile}
\end{figure}

The JPL-DE200 solar system ephemeris is adopted to convert the arrival time of photons to the barycenter of the solar system. After the solar barycenter corrections, the epoch-folding method \citep{1987A&A...180..275L} is used to search for the frequencies and to obtain the pulse profiles and periodic signals of the Crab pulsar. The total pulse profiles for the Crab pulsar observed by \textit{Insight-HXMT} in three energy bands (LE:1--10\,keV, ME:10--30\,keV, HE:27--250\,keV) are shown in Figure \ref{FIG:CrabProfile}. The counts are normalized by scaling the counts of the main peak to unity and shifting the mean value of the background phase (0.6--0.8) to zero. The Crab's pulse profiles have high signal-to-noise ratio due to the pointing observation mode and the large effective areas of the payloads on-board \textit{Insight-HXMT}. The main pulse of LE 
is delayed by \SI{864.7}{\micro\second} compared to \textit{NICER}/XTI. 
The accurate offset of LE is measured by the method in section \ref{subsec:offset}. The detailed processes of instrumental delay are described in Appendix \ref{sec:app}. 

\begin{figure}
	\centering
		\includegraphics[width=0.8\textwidth]{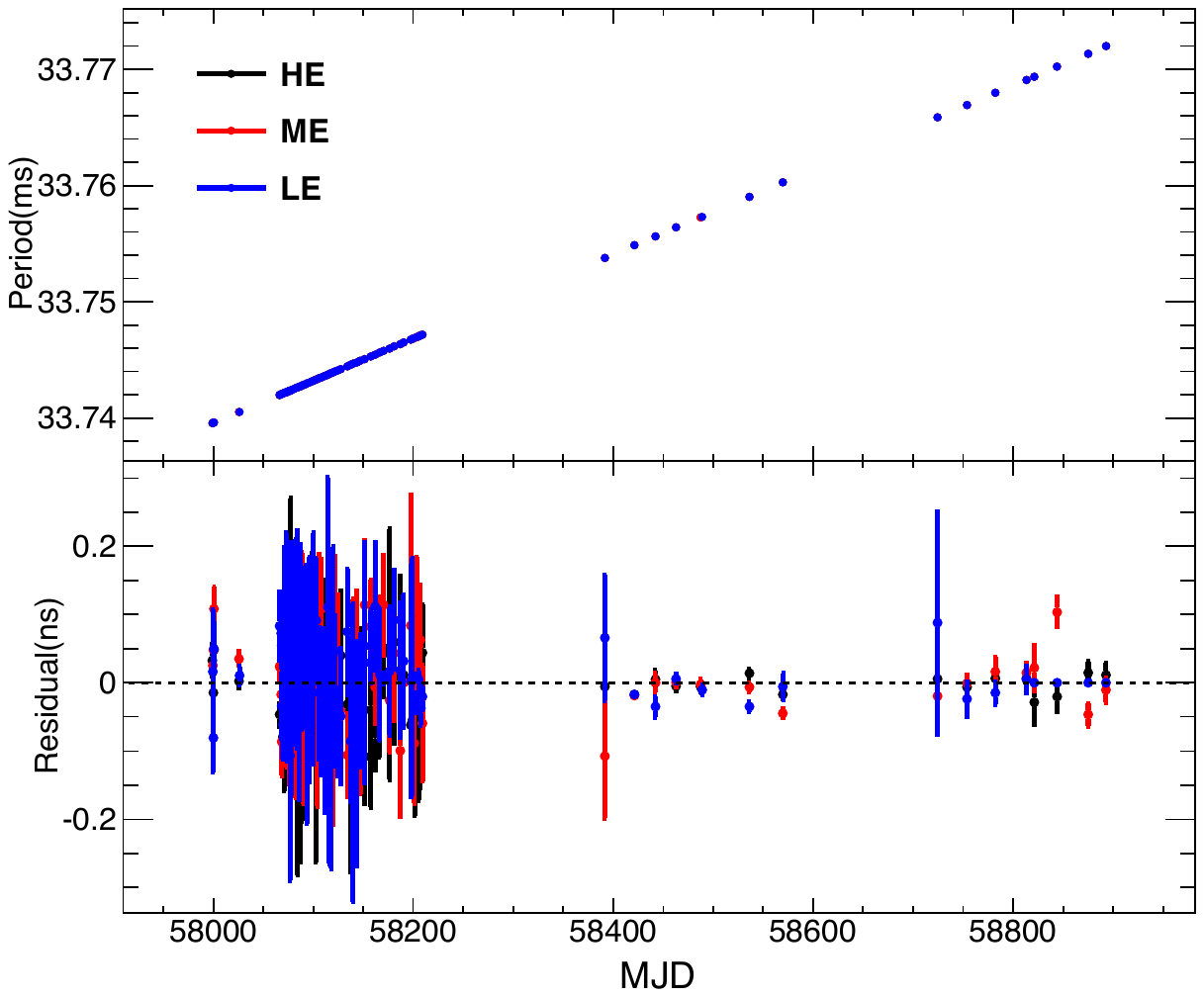}
	\caption{The top panel shows the periods of the Crab pulsar obtained by \textit{Insight-HXMT} on different dates. The bottom panel shows the difference between our results and those from radio measurements at the Jodrell Bank observatory.}
	\label{FIG:CrabPeridvsJB}
\end{figure}

The periods of the Crab pulsar measured on these 78 observations are plotted in Figure \ref{FIG:CrabPeridvsJB}. The well-known spin-down trend of the Crab pulsar can be revealed and the period agrees with the radio measurements \citep{1993MNRAS.265.1003L} at the Jodrell Bank observatory within 0.1\,ns.

\subsection{Offset and systematic errors of \textit{Insight-HXMT} timing system}
\label{subsec:offset}
The offset and the systematic errors of \textit{Insight-HXMT}'s timing system can be revealed by measuring the evolution of the absolute phase of the main peak over time. 
At first we investigate the systematic error using the Jodrell Bank radio ephemeris, similar to  \textit{INTEGRAL}/SPI \citep{2003A&A...411L..31K, molkov2010absolute}, and \textit{RXTE}/PCA \citep{2004Rots}. The systematic error of \textit{Insight-HXMT} timing system is within \SI{45}{\micro\second}. Similar results have been reported in \textit{RXTE} observations, where the systematic error due to the interstellar scattering and instrumental calibration of radio telescope is dominant \citep{2004Rots, 2012ApJS..199...32G}. Because this systematic error is relatively high, it is not precise enough to calibrate the timing system of \textit{Insight-HXMT} by directly using the radio ephemeris. Therefore, we use the quasi-simultaneous observations on the Crab with \textit{NICER} to produce the Crab ephemeris and then estimate the offset and systematic errors of \textit{Insight-HXMT}, given the good timing performance of \textit{NICER}/XTI \citep{2021Sci...372..187E}. 

The procedures for generating Crab pulsar's ephemeris using \textit{NICER}/XTI observations are described in the following. We redo the barycentric correction using JPL-DE405 solar system ephemeris for the data of quasi-simultaneous \textit{NICER} and \textit{Insight-HXMT} Crab observations. The spinning frequency and the pulse profile from each \textit{NICER} observation are obtained by the frequency search procedure mentioned in section \ref{subsec:search}. ToAs and their errors are calculated based on the pulse profiles using the fitting method mentioned in section \ref{subsec:toas}. The timing parameters are obtained by fitting the evolution of ToAs over time using \texttt{TEMPO2} \citep{2006tempo2}. This set of timing parameters is then used to epoch-fold the pulse profile and calculate a new set of ToAs. By iterate the above procedure of fitting and calculating ToAs, the timing parameters and ToAs converge to stable values. The ephemeris containing the timing parameters obtained by \textit{NICER}/XTI data is listed in Table \ref{tab:nicer_eph}.

\begin{table}[htb]\footnotesize
    \centering
    \begin{tabular}{c c c }
       \hline
           \textbf{Parameter} & \textbf{Time range 1} & \textbf{Time range 2} \\
           \hline
           RA (J2000) & \multicolumn{2}{c}{ $05^{\rm{h}}34^{\rm{m}}31^{\rm{s}}.973$ } \\
       DEC (J2000) & \multicolumn{2}{c}{$22^{\circ}00'52''.069$}  \\

       START (MJD) & 58064.99 &  58070.99\\
       STOP (MJD) & 58071.01 &   58078.01\\
       $t_0$ (MJD;TDB) & 58068 & 58080\\
       $f_0$ (Hz) & 29.63662156(2)  & 29.63623713(1)\\
       $f_1$ ($10^{-10}\,\rm{Hz}\,\rm{s}^{-1}$) & -3.701(4) & -3.704(1) \\
       $f_2$ ($10^{-18}\,\rm{Hz}\,\rm{s}^{-2}$) & -7(4).252 & 2(1).382 \\
       $f_3$ ($10^{-23}\,\rm{Hz}\,\rm{s}^{-3}$) & 5(4).431 & 0.5(2)02 \\
        \hline
    \end{tabular}
    \caption{Crab ephemeris fitted by \textit{NICER} observations which are simultaneously with \textit{Insight-HXMT}.}
    \label{tab:nicer_eph}
\end{table}

\begin{figure}
	\centering
		\includegraphics[width=0.9\textwidth]{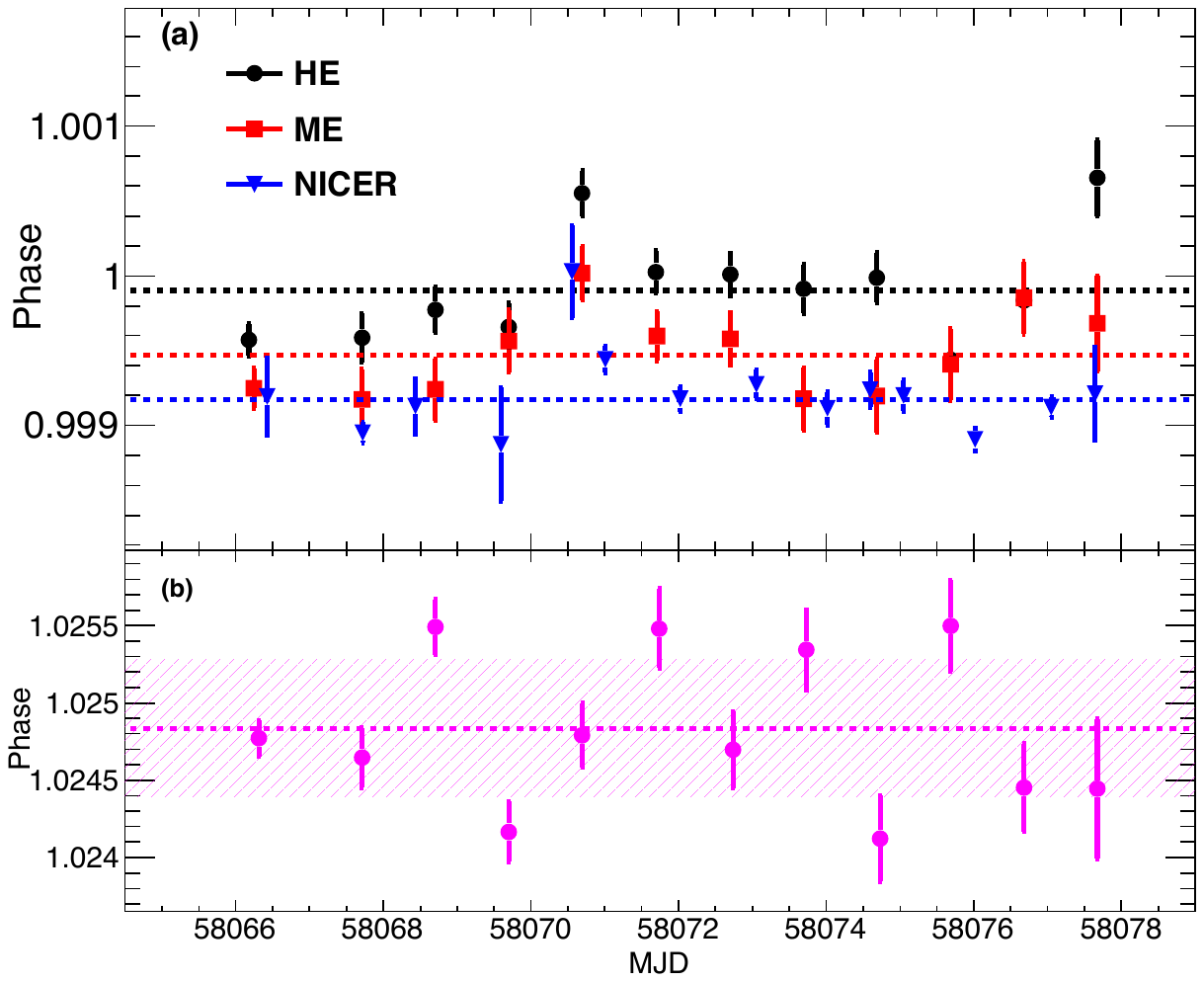}
	\caption{The top panel shows the best-fit values of the main pulse positions of the Crab pulsar using the timing ephemerides of \textit{NICER}. The black points, red squares, and blue triangles are the results for HE, ME, and \textit{NICER} respectively. The corresponding dotted lines are the mean phases for them. The bottom panel shows the best-fit values of the main pulse positions of Crab pulsar for LE. The purple dotted line is the mean phase for LE. The purple slashed area show the 1$\sigma$ (systematic error) confidence intervals for the mean phase.}
	\label{FIG:Crababsolutephase}
\end{figure}

Since the timing parameters of the Crab pulsar have been obtained from \textit{NICER} observations, the positions of the main peak observed by \textit{Insight-HXMT} can be measured from the epoch-folded pulse profiles. As shown in Figure \ref{FIG:Crababsolutephase}, the main peak position of each observation are plotted in different colors for LE (in purple), ME (in red), HE (in black), and \textit{NICER}/XTI (in blue). The dashed lines are the weighted mean positions of each payload respectively.

We define the offset as the deviation of the observed main peak position that observed by \textit{NICER}. The offsets of \textit{Insight-HXMT} are displayed in Table \ref{tab:errorandoffset}. The systematic errors and mean phase can be derived using the same method from equation (11) in \cite{2020JHEAp..27...64L} and the results are also displayed in Table \ref{tab:errorandoffset}. 

\begin{table}[htb]\footnotesize
  \centering
  \caption{The timing errors of \textit{Insight-HXMT} payloads respect to \textit{NICER}.}
  \label{tab:errorandoffset}
   \begin{tabular}{ccccc}
    \hline
     Payloads & mean  & systematic  & statistic       & offset   \\
             & phase  & error (\si{\micro\second}) &  error (\si{\micro\second})  & (\si{\micro\second}) \\
            \hline
      \textit{NICER} & 0.999172  & 5.5  & 0.9 &  0 \\
     HE & 0.999904 & 12.1 & 1.5  &  24.7\\
    ME &  0.999472 &  8.6 & 1.9  & 10.1 \\
    LE &  1.02483  & 15.8 &  2.1   &   864.7 \\
    \hline
    \end{tabular}
\end{table}

\subsection{Delay evolution with energy}

The deviation of the main peak obtained by ME and HE respect to \textit{NICER}/XTI observation is caused mainly by the intrinsic property of the Crab pulsar, because the instrumental time delay is about \SI{0.334}{\micro\second} and \SI{0.156}{\micro\second} for ME and HE, respectively. The instrumental delays estimated by the electronic design are described in Appendix \ref{sec:app}. 

\begin{figure}[H]
    \centering
    \includegraphics[width=0.8\textwidth]{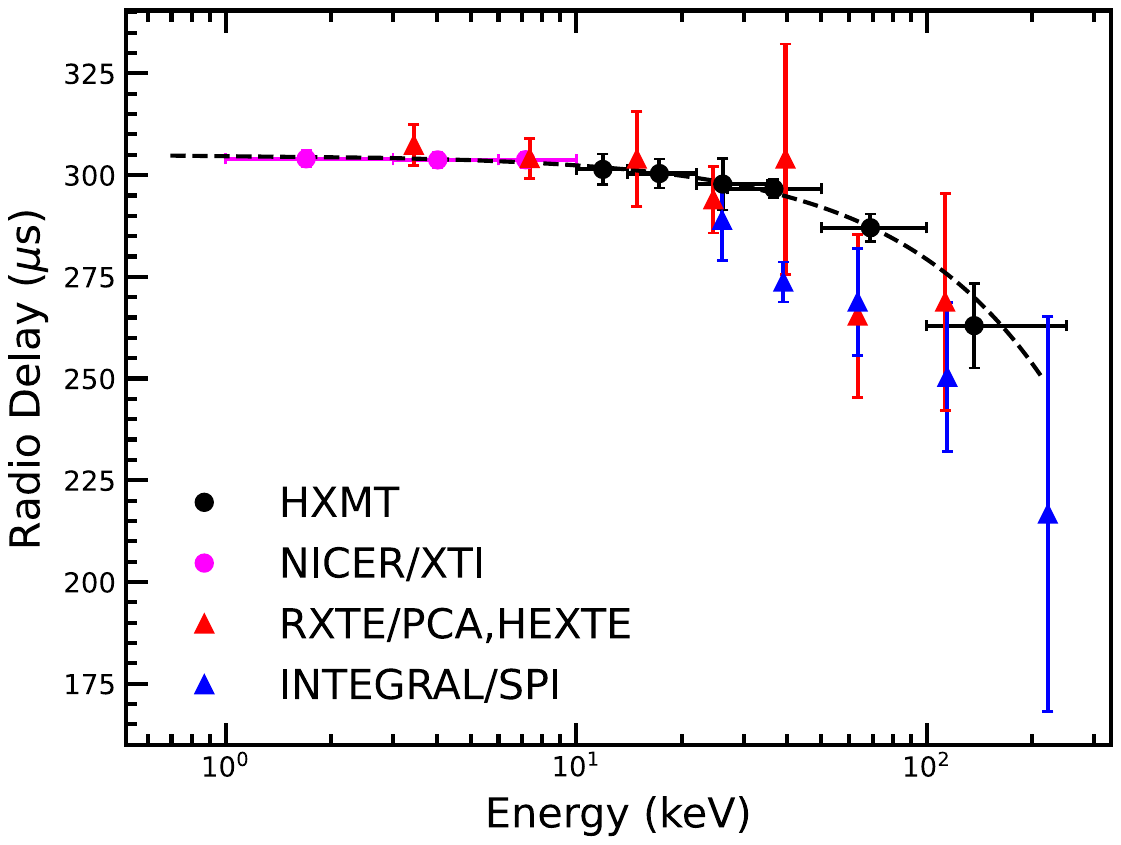}
    \caption{Radio delay of the main peak position for the Crab pulsar compared to the X-ray at different energies. The main peak delays of \textit{Insight-HXMT}/HE and \textit{Insight-HXMT}/ME with respect to \textit{NICER}/XTI are calculated. Considering the radio delay of \SI{304}{\micro\second} compared to XTI reported in \cite{2021Sci...372..187E}, we obtain the radio delay as a function of X-ray energy. The data are collected from \textit{NICER} and \textit{Insight-HXMT}, covering 1--250\,keV. The black dashed line is the best-fit to the data points of HE, ME, and XTI with a linear function. The purple \textit{NICER} and black \textit{Insight-HXMT} results are from this work. The radio delays of \textit{RXTE}/PCA, HEXTE (red triangles) and \textit{INTEGRAL}/SPI (blue triangles) reported in \cite{molkov2010absolute} are shifted downward by about \SI{10}{\micro\second} for comparison.}
    \label{fig:energyresolveddelay}
\end{figure}

The broad band observations on the pulsar indicate that the main peak phase of the Crab pulsar changes with energy \citep{molkov2010absolute}. The energy resolved phase of HXMT delay to the \textit{NICER} main peak can be determined as previously described. We split XTI, ME, and HE data in the first duration mentioned in Table \ref{tab:nicer_eph} into three energy bands, and get the position of the main peak in each energy band by Gaussian fitting. Figure \ref{fig:energyresolveddelay} shows the evolution of the peak position versus energy with respect to the 1--3\,keV peak position obtained from \textit{NICER}/XTI. We select the main peak position of \textit{NICER} XTI data in 1--3\,keV as reference position and calculate the delay of the main peak of difference profiles with different energy, as shown in Figure  \ref{fig:energyresolveddelay}. 
In order to compare with the results reported in \cite{molkov2010absolute}, a \SI{304}{\micro\second} radio delay compared to \textit{NICER}/XTI \citep{2021Sci...372..187E} is added to the offsets measured. Since the relative offsets of \textit{Insight-HXMT} compared with \textit{NICER}/XTI for 1--3\,keV have already been known, the radio delay relative to HXMT can also be derived as shown in Figure \ref{fig:energyresolveddelay}.

The delay decreases with energy at a rate of $0.26\pm0.02$\,\si{\micro\second\per\keV} by a linear fitting using the NICER and HXMT data. The radio delays of \textit{RXTE}/PCA, HEXTE and \textit{INTEGRAL}/SPI reported in \citep{molkov2010absolute} are also shifted downward by about \SI{10}{\micro\second} and shown in Figure \ref{fig:energyresolveddelay}. The results of RXTE are generally consistent with the linear trend, but the results from \textit{INTEGRAL}/SPI with large uncertainties is lower than the linear trend.
This rate is smaller than the value reported in \cite{molkov2010absolute}  $0.6 \pm 0.2$\,\si{\micro\second}, but consistent within $2\sigma$.

\section{Conclusions}
\label{sec:conclusion}
We have investigated the stability and accuracy of the timing system of \textit{Insight-HXMT} with the observation of the Crab pulsar. We find that the systematic errors of HE, ME, and LE are \SI{12.1}{\micro\second}, \SI{8.6}{\micro\second}, and \SI{15.8}{\micro\second}, respectively. The offsets obtained from the ToAs of Crab pulsar with respect to the \textit{NICER} measurements are HE: \SI{24.7}{\micro\second}, and ME: \SI{10.1}{\micro\second}. The non-physical offset of LE to \textit{NICER} is  ~\SI{864.7}{\micro\second} due to the special readout mechanism. 

Jointly using the \textit{NICER} and \textit{Insight-HXMT} data, the evolution of Crab pulsar's main peak position can be well fitted by a linear function of  $0.26\pm0.02$\,\si{\micro\second\per\keV}, which is smaller than but consistent with the results of \cite{molkov2010absolute} within $2\sigma$ confidence range. 

\appendix

\section{Internal instrumental delay}
\label{sec:app}
It is essential to provide the instrumental delay of the photons for studying the energy-dependence of the Crab main pulse and comparing the results with other instruments. The time system is corrected by the PPS signal, so the offset of the X-ray photons recording time related to PPS signal represents the instrumental delay. The total instrumental delay can be estimated by accumulating the typical response time of each electronic device in the electronic design.

\begin{figure}[H]
    \centering
    \includegraphics[width=1.0\textwidth]{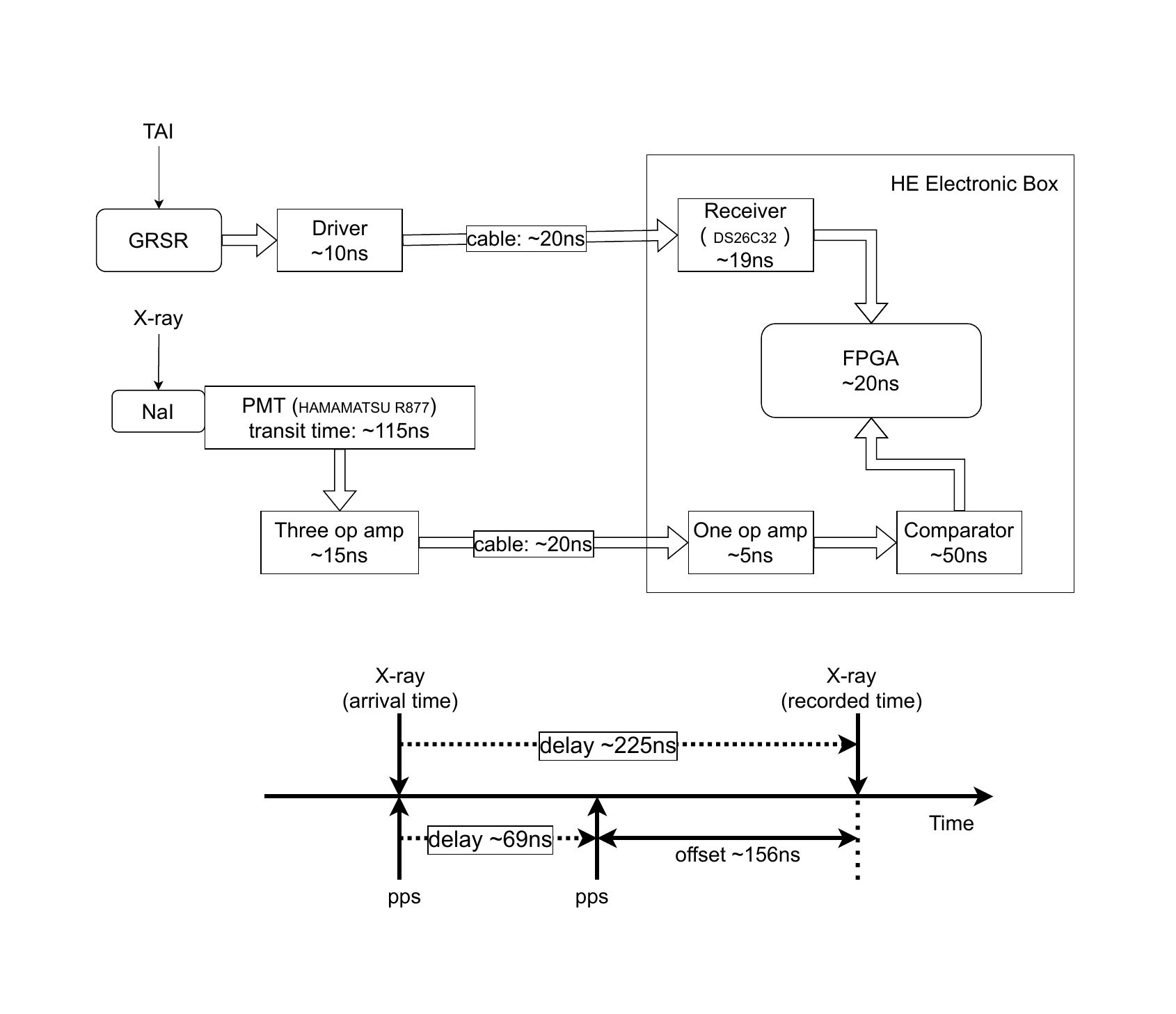}
    \caption{The upper diagram is the electronic design of the \textit{Insight-HXMT}/HE, and the lower diagram indicates the internal instrumental delay of detected X-ray photons. The PPS signal and X-ray photons are recorded by the FPGA through different electronic processes. For the PPS signal, the GPS receiver (GPSR) receives the international atomic time (TAI) signal and then passes through a driver that produces a \SI{10}{\nano\second} delay. The PPS then travels at a speed of \SI{5}{\nano\second\per\meter} over a cable of about \SI{4}{\meter} to the HE electronic box. One receiver spends \SI{19}{\nano\second} in signal processing. The FPGA sampling process generates another \SI{20}{\nano\second} delay. For X-ray photons, after depositing in NaI crystal, the photo-electrons are generated by the PMT and the transit time of PMT produces a delay of \SI{115}{\nano\second}. After three operational amplifiers (generating a delay of \SI{15}{\nano\second}), and through about \SI{4}{\meter} of cable (about \SI{20}{\nano\second} delay), it reaches the HE electronic box. Through another operational amplifier (about \SI{5}{\nano\second}) and a comparator (generating a delay of \SI{50}{\nano\second}), it arrives at the FPGA and is recorded. A \SI{20}{\nano\second} of sampling time is also required for the FPGA to process the X-ray photons. The lower diagram indicate that the X-ray photons have approximately \SI{156}{\nano\second} offset to the GPS PPS signal.}
    \label{fig:HEelectronicdesign}
\end{figure}

\begin{figure}[H]
    \centering
    \includegraphics[width=1.0\textwidth]{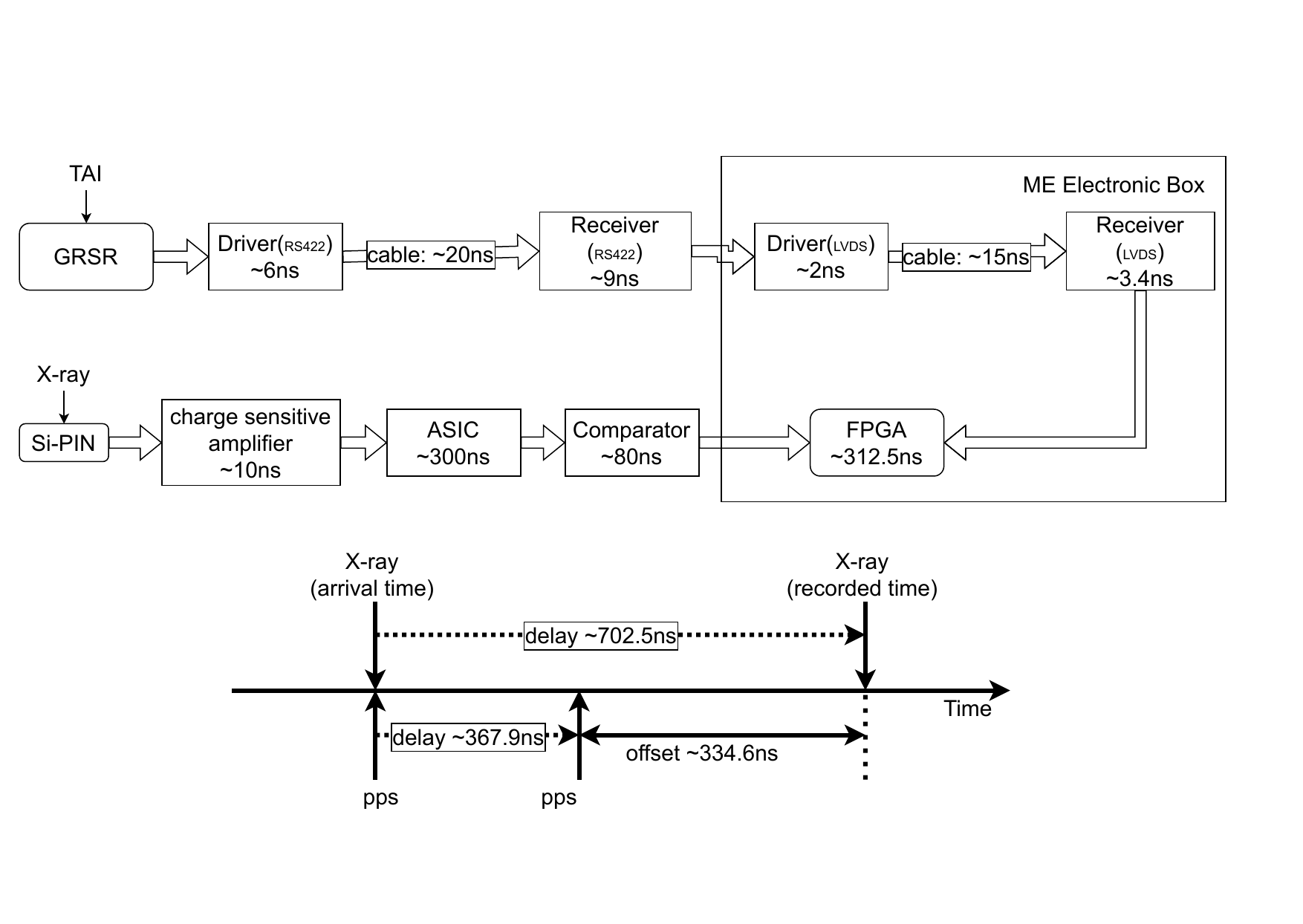}
    \caption{The upper diagram is the electronic design of the \textit{Insight-HXMT}/ME, and the lower diagram indicates the internal instrumental delay of detected X-ray photons. For the PPS signal, the GPS receiver (GPSR) receives the international atomic time (TAI) signal and then passes through a driver that produces a \SI{6}{\nano\second} delay. The PPS then travels at a speed of \SI{5}{\nano\second\per\meter} over a cable of about \SI{4}{\meter}. Then it is processed by a receiver, which generates a \SI{9}{\nano\second} delay. In the ME electronic box, a signal passes through another driver (costs \SI{2}{\nano\second}), a \SI{3}{\meter} cable (costs \SI{15}{\nano\second}), and another receiver (costs \SI{3.4}{\nano\second}) and is recorded by the FPGA. The FPGA sampling process generates a \SI{312.5}{\nano\second} delay. The total delay of PPS signal is about \SI{367.9}{\nano\second}, as indicated in the lower diagram. For X-ray photons, they are detected by Si-PIN detector. The charge sensitive amplifier processes the signal and produces a \SI{10}{\nano\second} delay. Through the ASIC (\SI{300}{\nano\second}) and a comparator (\SI{80}{\nano\second} delay), it recorded by the FPGA in the ME electronic box. The total delay of an X-ray photon is about \SI{702.5}{\nano\second}, as indicated in the lower diagram. The offset of X-ray photons compared to PPS is about \SI{334.6}{\nano\second}.}
    \label{fig:MEelectronicdesign}
\end{figure}

For \textit{Insight-HXMT}/HE and \textit{Insight-HXMT}/ME, the electronic design are shown in Figure \ref{fig:HEelectronicdesign} and Figure \ref{fig:MEelectronicdesign}. The delays are \SI{0.156}{\micro\second} and \SI{0.335}{\micro\second}, respectively. The instrumental delays for HE and ME are negligible, compared to the energy-dependent evolution of \SI{0.26}{\micro\second\per\keV} as discussed in \ref{sec:Results}.

\begin{figure}[H]
    \centering
    \includegraphics[width=1.0\textwidth]{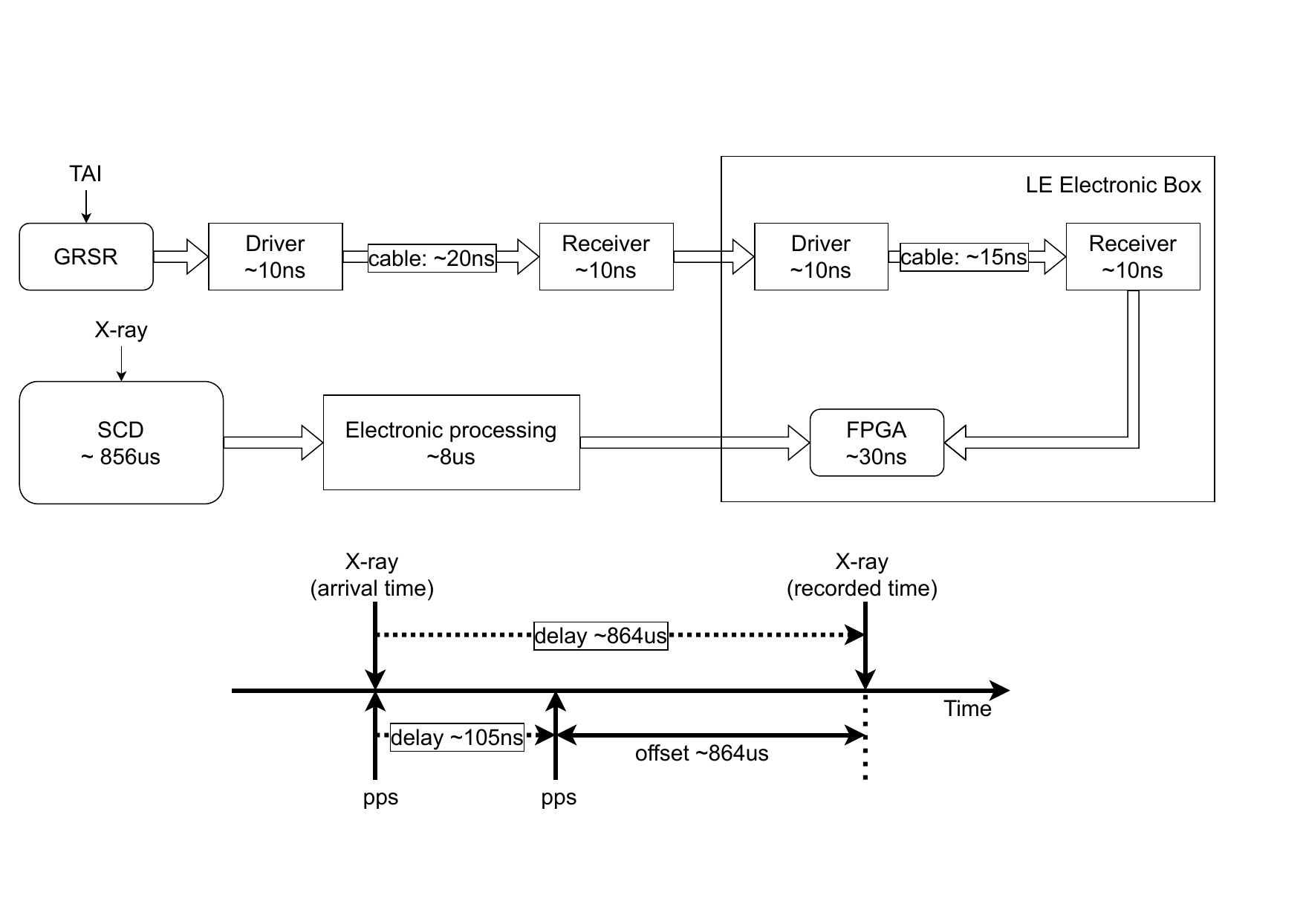}
    \caption{The upper diagram is the electronic design of the \textit{Insight-HXMT}/LE, and the lower diagram indicates the internal instrumental delay of detected X-ray photons. For the PPS signal, the electronic processes are similar to ME design. The GPS receiver (GPSR) receives the international atomic time (TAI) signal and then passes through a driver that produces a \SI{10}{\nano\second} delay. The PPS then travels at a speed of \SI{5}{\nano\second\per\meter} over a cable of about \SI{4}{\meter}. Then it is processed by a receiver, which generates a \SI{10}{\nano\second} delay. In the LE electronic box, a signal passes through another driver, a \SI{3}{\meter} cable, and another receiver and is recorded by the FPGA. The FPGA sampling process generates a \SI{30}{\nano\second} delay. The total delay of PPS signal is about \SI{105}{\nano\second}, as indicated in the lower diagram. For X-ray photons detected by LE SCD detectors, the delay is mainly contributed by the special read-out mechanism. The main pulse of the Crab pulsar will b delayed by \SI{856}{\micro\second}. The total electronic processes produce about \SI{8}{\micro\second} delay. Thus the instrumental offset of X-ray photons of LE is about \SI{864}{\micro\second}, as indicated in the lower diagram. }
    \label{fig:LEelectronicdesign}
\end{figure}

The electronic design of \textit{Insight-HXMT}/LE is shown in Figure \ref{fig:LEelectronicdesign}, in which the total delay is \SI{864}{\micro\second}. The delay of LE is mainly due to the special readout mechanism of the swept charge devices for LE \citep{2019JHEAp..23...23Z, 2021RAA....21....5Z}. We note that \cite{2019JHEAp..23...23Z} reported the probability distribution of the delay time, in which the distributions are measured both in-orbit  and on-ground. The delay probability distribution is very different between the pre-launch and the in-orbit measurements, because there were no collimators installed during the on-ground measurement. 
To get the delay of this readout mechanism of LE, the pulse profile of the Crab pulsar measured by NICER is convolved with the delay distribution measured in-orbit. The peak of the main pulse is delayed by \SI{856}{\micro\second} to the real pulse profile.
The electronic process also takes about \SI{8}{\micro\second} delay before it sends the signal to the FPGA of LE box as shown in Figure \ref{fig:LEelectronicdesign}.
Therefore, the total delay of LE is about  \SI{864}{\micro\second}, which is similar to the measured phase shift of the Crab pulsar compared with NICER.

\section*{Acknowledgements}
   This work is supported by the National Natural Science Foundation of China under grants (No. U1838105, U1838201, U1838202, U1938108, U1938102, U1938109, U1838104) and the National Program on Key Research and Development Project (Grant No.2016YFA0400801). This work made use of data from the \textit{Insight-HXMT} mission, a project funded by China National Space Administration (CNSA) and the Chinese Academy of Sciences (CAS).



\bibliography{HXMTTimingcal}{}
\bibliographystyle{aasjournal}

\end{document}